\documentclass[12pt,preprint]{aastex}
\usepackage{verbatim}
\begin{document}

\title{Newly Determined Explosion Center of Tycho's Supernova and the Implications for Proposed Ex-Companion Stars of the Progenitor}
\author{Zhichao Xue \& Bradley E.  Schaefer\affil{Physics and Astronomy, Louisiana State University, Baton Rouge, LA 70803}}

\begin{abstract} 
`Star G', near the center of the supernova remnant of Tycho's SN1572, has been claimed to be the ex-companion star of the exploding white dwarf, thus pointing to the progenitor being like a recurrent nova.  This claim has been controversial, but there have been no confident proofs or disproofs.   Previously, no has seriously addressed the question as to the exact explosion site in 1572.  We now provide accurate measures of the supernova position by two radically different methods.  Our first method is to use the 42 measured angular distances between the supernova in 1572 and bright nearby stars, with individual measures being as good as 84 arc-seconds, and all resulting in a position with a 1-$\sigma$ error radius of 39 arc-seconds (including systematic uncertainties).  Our second method is to use a detailed and realistic expansion model for 19 positions around the edge of the remnant, where the swept-up material has measured densities, and we determine the center of expansion with a chi-square fit to the 19 measured radii and velocities.  This method has a 1-$\sigma$ error radius of 7.5 arc-seconds.  Both measures are substantially offset from the geometric center, and both agree closely, proving that neither has any significant systematic errors.  Our final combined position for the site of the 1572 explosion is J2000 $\alpha$=0h 25m 15.36s,  $\delta=64^{\circ} 8' 40.2"$, with a 7.3 arc-second 1-sigma uncertainty.  Star G is rejected at the 8.2-$\sigma$ confidence level.  Our new position lies mostly outside the region previously searched for ex-companion stars.

\end{abstract}

\section{Introduction}
The nature of the progenitor system for a supernova (SN) of Type Ia (SNIa) is among the most important questions in astrophysics (Ruiz-Lapuente 2014).  This progenitor problem has been controversial for decades.  While the empirical application of SNIa as cosmological indicator has successfully led to the discovery of the the accelerating Universe (Perlmutter et al.  1999; Riess et al.  1998), we still need to close the gap in theoretical understanding of them.  A SNIa is known to be the product of a close binary system where one of the stars is a carbon/oxgyen white dwarf (CO WD).  The companion could be a main-sequence star, sub-giant star,  red giant star or another CO WD.  SNIa are produced by one of two channels: the single-degenerate (SD) channel (Whelan \& Iben 1973; Iben \& Tutukov 1984) and the double-degenerate (DD) channel (Webbink 1984).  The SD scenario has an ordinary companion star spilling matter onto the WD, accumulating material until the WD gets near the Chandrasekhar mass, thus initiating a thermonuclear explosion.  Within the SD scenario, the progenitor might have been a recurrent nova, a symbiotic star, a persistent super-soft X-ray source, or a helium star.  In all cases, SD progenitors will leave behind a luminous ex-companion star, now orbiting nothing, that has been battered by the supernova blast.  The DD scenario has two CO WDs in close orbit, in-spiraling until the two stars merge, with a combined mass near the Chandrasekhar limit, thus triggering a thermonuclear explosion that leaves nothing but a beautiful expanding remnant.  

One of the most direct and effective ways to distinguish between progenitor models is to search for a possible ex-companion within a supernova remnant (SNR), as first proposed by Ruiz-Lapuente (1997).  Her idea is that the different models predict different surviving companion stars, for example, a symbiotic progenitor would leave behind a red giant star, a recurrent nova would likely leave behind a red giant or a sub-giant star, a super-soft source would leave behind a sub-giant or a massive main sequence star, while a DD progenitor would leave behind no companion.  This method can starkly distinguish between the SD and DD scenarios simply by looking for whether there is a luminous ex-companion star or not.  This method was first turned towards Tycho's supernova of 1572, which is now confidently known to have been a SNIa event from observed light echos (Rest et al.  2008; Krause et al.  2008).  Ruiz-Lapuente et al.  (2004, RL04) searched deep within the central area of Tycho's SNR and found a G-type sub-giant (labeled star `G') with high proper motion and probably the right distance to be an ex-companion.  RL04 noted that the ex-companion is like the secondary star in the recurrent nova U Sco, thus pointing to recurrent nova as being the SNIa progenitor.  Our group at the Louisiana State University applied the Ruiz-Lapuente method by using the {\it Hubble Space Telescope}  ({\it HST}) to look deep in the center of SNRs in the Large Magellanic Cloud known to be SNIa by the spectrum of their light echoes.  The most startling case was for SNR 0509-67.5, where the 3-sigma uncertainty circle in the center was completely empty of any point source to V=26.9 (corresponding to $M_V=8.4$), pointing to the DD model because all published SD models were strongly rejected (Schaefer \& Pagnotta 2012).  (There is a background z=0.031 galaxy in the central error circle; Pagnotta et al.  2014.) Edwards et al.  (2012) applied the same method to SNR 0519-69.0, and demonstrated that there are no giant or sub-giant ex-companions, ruling out most SD models.  Pagnotta \& Schaefer (2015) have looked in the centers of SNR 0505-67.9 and SNR 0509-68.7 to find many red giants and sub-giants, so these two SNRs are not useful.  Further work has demonstrated that SN 1006 has no ex-companions going deep (Gonzalez Hernandez et al.  2012; Kerzendorf et al.  2012), and that Kepler's SN (SN 1604) has no red-giant ex-companions, nor any sub-giant ex-companions down to 10 $L_{\odot}$ (Kerzendorf et al.  2014).  In all, we have a stark case that five SNIa SNRs do not have the SD-predicted ex-companions, with one to very deep limits.  The one possible exception is star G in Tycho's SNR.

If Tycho star G is indeed the ex-companion, then we have immediately eliminated the DD scenario and the symbiotic model, forcing ourselves into a SD answer, at least for this one supernova.  Following the publication of RL04, our community expressed substantial amounts of skepticism.  Based on many comments heard and read at the time, most of the initial skepticism had poor bases; either the reaction was that the progenitor systems are DD so star G is not the ex-companion (assuming that which is trying to be proven) or that the proper motion of star G is perpendicular to the center of the SNR (but the position of star G in 1572 was still inside the RL04 error circle for the site of the explosion, so all is OK).  Despite these primitive reactions, various follow-up investigations of star G have turned up good reasons to be critical of the claim that it is an ex-companion star.  A surviving companion of a SNIa should show some distinctive features.  First, it should show up close to the explosion site of the supernova.  Second, it might show peculiar high velocity compared to the average motion of other local stars, due to the now-unbounded orbital velocity of the previous binary system.  Third, the stellar atmosphere should be contaminated by accreted SN ejecta, which mainly consists of heavy elements like iron and nickel.  Thus, heavy metal lines should show up in the spectrum of the proposed ex-companion star.  Each feature has been extensively debated, with papers and arguments going several levels deep.  For the issue of the high proper motion, Kerzendorf et al.  (2009) first challenged the claim of high proper motion in RL04, but later Kerzendorf et al.  (2013) and Bedin et al. (2014) confirmed the RL04 detection of high velocity of star G with high precision.  Kerzendorf et al.  (2009) raised the issue that a tidally synchronized companion star would lead to a fast rotating ex-companion star, while they measured that star G is slow rotating.  Gonzalez-Hernandez et al.  (2009) pointed out this can be well explained in part by the SN explosion blasting away the outer layers of the companion, hence carrying away angular momentum, and in part by the expansion of the ex-companion back to an equilibrium state and slowing its rotation as it expands.  Further, Pan et al.  (2012) and Liu et al.  (2013) suggested that the rotational velocity would be reduced due to the impact of SN ejecta.  Gonzalez Hernandez et al.  (2009) have claimed that star G has an anomalously high nickel abundance, with this being disputed by Kerzendorf et al.  (2012), while Bedin et al.  (2014) re-evaluated the nickel abundance to be anomalously high.  The argument that Star G is a giant star with distance $\sim$10 kpc is brought up by Schmidt et al.  (2007).  But the high resolution spectra obtained with Keck again confirmed Star G is a sub-giant with distance that is compatible with being inside the SNR ($\sim$3 kpc).  In the absence of any decisive argument, workers have proposed that Tycho star `B' or Tycho star `E' might be the long-sought ex-companion of SN 1572 (Kerzendorf et al.  2013; Ihara et al.  2007).  Our judgment of all this contradictory evidence is that no persuasive case has been made that any star either is or is-not an ex-companion.

We have realized that there is one critical question that no group has seriously addressed, and that is to the position in the sky of the original supernova explosion.  Various groups had reported geometric centers for the SNR, all apparently with only casual care, and RL04 took some sort of an average, then selected a reasonable error circle radius (39") as the central 15\% of the SNR radius.  We further realized that the position of the original supernova explosion can be greatly improved; and we have two completely independent methods to do this.  The first method is to use the original astrometry from 1572, where Tycho Brahe himself and six other observers reported 42 angular distances between nearby bright stars and the supernova itself, with accuracies measured to be as good as 84".  A chi-square fit to all these observations produces a position of the supernova with a 1-$\sigma$ error ellipse 28" by 35", all with no uncertainties from expansion, distance, proper motion, or extinction.  The second method is to construct an analytic expansion model for the SNR, where we reproduce the observed expansion velocities and angular sizes for 19 positions around the edge of the SNR shell for which the densities of the swept up material have been measured previously.  Our chi-square fit reproduces the slightly out-of-round shape of the SNR, and shows that the relatively high density of swept up material to the northeast and northwest have made the apparent geometric center of the SNR offset to the south-southeast of the original explosion site.  Our position of the explosion site has a 1-$\sigma$ accuracy of 7.5".  In this paper, we will present our two measures of the original supernova position, and this will provide a simple and convincing resolution of the conflict as to whether Tycho star G can be the ex-companion or not.

\section{Explosion Position From Astrometry in 1572}

The position of the explosion site of Tycho's supernova can be measured to surprisingly good accuracy as based on astrometric measures of the position of the bright supernova in the year 1572 (and 1573).  Seven different observers report 42 positions, mostly as angular separations between nearby bright stars and the supernova.  For Tycho's observations, the one-sigma error as determined from his reported distances between bright stars is 1.4 arc-minutes.  Simplistically, we could expect a combined uncertainty in one direction of $1.4'/(0.5\times42)^{0.5}$ or 0.3 arc-minutes, with this being adequate to distinguish whether Star G is at the explosion site.

The celestial position of Tycho's supernova has been derived by many prior workers (Hind 1861; Argelander 1864; Bohme 1937; Baade 1945; Stephenson \& Clark 1977; Clark \& Stephenson 1977; Green 2004), with the positions summarized in Table 5 of Green (2004).  These works have been aimed at identifying the area of the sky to look for a SNR, as well as to understand the historical observations.  But different analyses are needed if we are addressing the astrophysics question ``What is the best position for the explosion site?", instead of questions of interest to historians.  Prior to Green (2004), essentially all attention was devoted to those observations of Tycho himself, and generally only using his three angular distances from $\alpha$, $\beta$, and $\gamma$ Cas.  In addition, no real error analysis is presented, with this being critical for evaluating whether the position of Star G is consistent with the explosion site.  Finally, each analyst used only a small fraction of the available measures, so their derived positions can be greatly improved.  In this paper, we will make substantial improvements on the prior work because we will use {\it all} observations with appropriate error bars, and we will construct real error ellipses for both statistical and systematic errors as taken from a standard chi-square analysis.  

\subsection{The Astrometry of the Supernova}

Tycho's observations of the supernova are so famous that they hide the fact that many other observers also provided detailed records, including measured positions of the visible supernova.  Green (2004) has collected, analyzed, and emphasized these additional observations, and indeed, it is from this paper that we became aware of their existence, quality, and number.  Seven European astronomers have their astrometry surviving, for a total of 42 positional measures of the supernova.  (Many more positional measures were reported, but these give positions that are useless for the task of deriving the best astrometric position of the exploding star, for example, because they report altitudes with uncertain times.)  Thirty-eight of these measures are simply reports of the angular distance between some bright star and the supernova, with these being made with `Jacob staff' (cross-staff) or sextant.  Four of these positional measures are reports that the supernova appeared exactly in line with two other stars, such that the three stars all fall on a great circle in the sky with the supernova in the middle.  These 42 observations are listed in Table 1.

Tycho reported on his full astrometry data set in his highly influential book {\it Astronomiae Instauratae Progymnasmata} in 1602, with this containing angular distances to the supernova from 12 nearby bright stars.  These supersede his earlier reported angles (in a 1573 publication), with the differences being that  he had gained much experience in appropriate corrections (e.g., the parallax associated with the sighting mechanism) and analysis of his old data.  Stephenson \& Clark (1977) report on their analysis of 36 measures by Tycho of the angular distance between two bright stars (for which the exact distance is now known) and find a one-sigma measurement error of 1.4 arc-minutes (84").  This contemporary measure provides the best estimate of the total error for Tycho.

In modern times, three attempts have been made to select or correct Tycho's distances.  First, Stephenson \& Clark (1977) have rejected the angular distances to Polaris and $\beta$ Cas solely because they were regarded as outliers.  However, our analysis shows them to be not extreme, and they certainly are far from being three-sigma outliers.  These two distances are valid information about the position of the supernova and hence should not be discarded, while they provide information about the real measurement uncertainties and must be included.  Second, Green (2004) has speculated that Tycho's reported distances to $\alpha$ Per and $\alpha$ Aur might have been calculated, with no basis other than that they appear only in the third part of the {\it Progymnasmata}.  But these two stars yield circles that are significantly far from any center (e.g., see the next paragraph), and so the case for calculation cannot be right, and we have included these two stars as valid measures.  Third, Stephenson \& Clark (1977) discarded two stars, ignored two other stars (leaving only 8 stars of Tycho's 12 with distances), looked at the residuals {\it from the center of the supernova remnant}, and concluded that Tycho's distances have a previously unknown systematic error where they are reported 2.6 arc-minutes too large.  With this, their derived position suddenly is fairly close to the center of the supernova remnant.  Such a set of procedures has no historical basis or support and is clearly just biasing the result towards some assumed position, and this must be rejected.  Also, such outliers and corrections do not appear in Tycho's contemporaneous observations of the angular distances between these same bright stars (Stephenson \& Clark 1977).  Such late-time corrections or selections are tempting, but they must be avoided at all costs because modern researchers do not have access to Tycho's raw data or his analysis procedures.

A possible problem arises because the reported angular distances from $\alpha$, $\beta$, and $\gamma$ Cas in Tycho's 1602 book all meet exactly (to within 2.5 arc-seconds) at one point on the sky.  Any two angular distances define two circles on the sky that meet, in general, at two points, while the probability that any third circle will touch within a few arc-seconds of one of these meeting points is small.  So how did it happen that just these three stars (and none other) coincide exactly to a unique point?  Earlier versions of Tycho's measures for these three stars had agreement at only the 2 arc-minute level.  So something in Tycho's corrections and re-reductions made the first three stars `perfect'.  Although we can now never know, likely this is all innocent with Tycho merely using these three stars to calibrate his corrections in some way.  Nevertheless, this alerts us to some level of systematic errors imposed by Tycho's analysis method.  Still, the contemporaneous measures of the distances between bright stars have all of Tycho's measurement and systematic errors , so their scatter (1.4 arc-minutes) should still be the best measure of Tycho's total one-sigma uncertainty.

Thaddeaus Hagecius ab Hayck observed the supernova from Prague and published six angular distances in 1574.  As part of the vigorous discussion amongst European astronomers concerning astrometric methods, inspired by the supernova, Hagecius corrected two of these angular distances and published these in a not-surviving revision of his book, although these corrected positions have been included in Tycho's {\it Progymnasmata}.

Jeronimo Munoz reports four angular distances from 1573 as made in Valencia.  Like all observers, he reported angular distances from $\alpha$, $\beta$, and $\gamma$ Cas, and these produce a consistent position that is about 6 arc-minutes from the modern supernova.  The fourth angular distance, from Polaris, produce a circle that only approaches to 25 arc-minutes from the supernova.  So this alerts us that the Munoz astrometry has uncertainties that are very large.

Cornelius Gemma published a book in 1575 that contained 9 angular distances as observed from Louvain.  Like Tycho, his originally published distances for $\alpha$, $\beta$, and $\gamma$ Cas in 1573 showed a reasonable consistency and position, while his final positions for the same three stars produced an exact coincidence to under one arc-second.  (The other stars disagree with this position by a median of 20 arc-minutes.)  So apparently, both Tycho and Gemma applied some corrections to their original distances that were somehow calibrated by these three stars.  The result is a set of distances that are only in poor agreement, and this means that the real accuracy of Gemma's observations are poor.

Bartholomew Reisacher, in 1573 from Vienna, made one distance measure, from $\kappa$ Cas (the nearest visible star to the supernovae).  With only one distance, Reisacher's data cannot produce a unique position on the sky, and it has been almost completely ignored.  However, when we are putting together all the astrometry, Reisacher's one observation must be included.

Thomas Digges, observing from London, reports on 6 angular distances from bright stars to the supernova.  He also reports on two pairs of stars, each of which define arcs along a great circle which contains the supernova.  That is, he tells us that the supernova was exactly on the great circle defined by the stars $\beta$ Cep and $\gamma$ Cas, as well as on the great circle defined by the stars $\iota$ Cep and $\delta$ Cas.  This observation was made by using a single thread, held taut in a holder, that can be moved around and rotated until it was precisely occulting the stars.  Green (2004) reports his practical experience with trying this alignment technique on steady bright stars.

Michael Maestlin, observing from Tubingen, also reported two pairs of stars whose great circles intersected at the supernova.  One of these star pairs ($\iota$ Cep and $\delta$ Cas) was identical to that reported by Digges, but the other ($\beta$ Cas and $\lambda$ UMa) is different.  It so happens that all three unique great circle arcs cross within 10 arc-seconds, and this implies good accuracy for this method, a bit of good luck, or some combination of luck and accuracy.  The accuracy of this method of alignments cannot be too poor, say, fractions of a degree, as otherwise it would be very unlikely to get an agreement to 10 arc-seconds.

\subsection{The Analysis Method}

Green (2004) has done the hard work of pulling out the astrometry data from the original sources from soon after 1572, and understanding their methods and conditions.  Green has also determined the modern names for the quoted comparison stars as well as calculated their astrometric positions for the equinox of 2000.0 and the epoch of 1573.0.  (That is, the proper motions of the bright stars are extended back to the time of the supernova, and these old positions are reported in the modern J2000 system.)  His results are the basis for our new derivation of the best position for the exploding supernova.

We use the standard method of chi-square, where we find the position on the sky that has the minimum total deviation (in units of the measurement uncertainty) between the position and the observations.  This approach has three strong advantages over previous analysis.  First, instead of some {\it ad hoc} method with unknown statistics, the chi-square method is universally known and understood.  Second, the chi-square method can easily handle any or all the observations, and it can easily be used to combine results from the reported distances and three-star alignments.  Third, the chi-square method produces real error bars, and indeed, these can be asymmetric error regions on the sky, with various known probability levels.  The result is a well-understood best-fit position on the sky as derived from {\it all} the astrometry measures and the real error region.

The bright stars and their angular distances ($\Theta$) from the supernova are tabulated in Table 1, while the one-sigma uncertainties in these distances ($\sigma$) will be given below.  Our chi-square procedure starts off with a trial position on the sky, for which we then calculate the angular distances between the trial position and all the stars ($\Theta_{trial}$).  The chi-square statistic will be simply the sum over all observations as $\chi^2 = \Sigma [(\Theta - \Theta_{trial})/\sigma]^2$.  The trial position is then varied around the sky until the minimum $\chi^2$ is reached ($\chi^2_{min}$), and that is the best-fit position.  The one-sigma error region will be the locus on the sky (roughly elliptical in shape) over which the chi-square equals $\chi^2_{min}+1$.  The three-sigma error region is the locus where the chi-square equals $\chi^2_{min}+9$.  We now have a method that produces the best-fit position (and its error bars) for any number of data points, and the calculations are all easy and fast.

The three-star alignments are handled by deriving a position on the sky that is exactly 90$^{\circ}$ from the two end stars, and then calculating the distance from this point to the trial supernova position.  The observation corresponds to this distance being 90$^{\circ}$.  With this formulation, these three-star alignments then function exactly like any other measured distance.

For the chi-square method, it is critical that we use reasonable values of the measurement uncertainty in the distances.  For Tycho's observations, we have a reliable measure (from his observations of distances between bright stars) that $\sigma_{Tycho}=1.4'=0.023\degr$.  For the other observers, we have effectively looked at the consistency of their reported angles as a measure of their real measurement errors.  For this, we have derived the best fit position for each observer individually, and looked at the RMS scatter in the deviations of the reported distances for this position.  Highly accurate observations will result in a small scatter of the deviations, while poor observations will feature large deviations and scatter.  In practice, we simply vary $\sigma$ until the best fit has a reduced chi-square equal to unity.  With this, we find $\sigma_{Hagecius}=0.153\degr$, $\sigma_{Munoz}=0.82\degr$, $\sigma_{Gemma}=0.38\degr$, and $\sigma_{Digges}=0.042\degr$.  (For the historical question of the relative accuracy of the various observers, we now have quantitative measures, with this being better than evaluating the apparent skills of the observer from their words on technique.  We now see that only Digges approached Tycho in his astonishing accuracy, while Munoz and Gemma were poor in accuracy.)  This method cannot be applied to Reisacher's one observation, so we have presumed to set his uncertainty to the geometric mean of the other five observers with angular distances, so $\sigma_{Reisacher}=0.136\degr$.

To make one evaluation of the uncertainty for the three-star alignments, we have performed a series of naked eye measures of three-star alignments by holding up a thread under a dark sky.  This recreation will not provide proof of any value, but it will be a valid indication of the appropriate size scale.  The procedure was simply to lie back under a dark and clear sky, pick a star as the middle for the three, and rotate the thread around the middle star until two stars appeared in approximately the same line.  With a candidate alignment, care was then taken to hold the eye and hands steady (by bracing the head and arms on a fencepost or the ground), using vision from only one eye, and carefully moving the thread until the stars are occulted by the thread.  Our experience is that we can tell whether the middle is exactly in the great circle arc or is slightly offset, so each of our alignments was graded as being either `exact' or (slightly) `offset'.  The next day, we then took the identified end stars, calculated a point on the sky that was 90$\degr$ from both, calculated the angle between this point and the middle star, and the difference from 90$\degr$ is taken to be the offset from a perfect great circle.  It is easy to find these three-star alignments, and we collected a total of 30.  We found a sharp dividing line at close to 0.2$\degr$, where deviations smaller than this were called `exact' and deviations larger than this have the middle star noticeably `offset'.  With possible alignments being evaluated as `exact' for deviations from roughly -0.2$\degr$ to +0.2$\degr$, the typical measurement uncertainty for any alignment reported to be `exact' is roughly $\pm0.1\degr$.  There is an additional source of uncertainty simply due to the fact that perfect alignments might not be available on the real sky, so the observer might have to settle for less than perfect alignments to report.  However, we found in practice that even regions with high galactic latitude will always have good three-star alignments, and Milky Way regions will have many alignments of high accuracy for the observer to choose between.  To give a high-latitude example, the star Arcturus ($\alpha$ Boo) is the middle star of great circle segments defined by $\theta$ CrB and $\gamma$ Vir, $\beta$ Boo and $\alpha$ Vir, and $\beta$ Her and $\eta$ Boo, with deviations of 0.11$\degr$, 0.18$\degr$, and 0.15$\degr$ respectively.  To give another example, for a star in the Milky Way region close to the supernova, the star $\gamma$ Cas is on the great circles between $\delta$ Cas and $\alpha$ Cyg, between $\zeta$ Cep and $\alpha$ Per, and between $\upsilon$ UMa and $\alpha$ Cas (with deviations of 0.14$\degr$, 0.10$\degr$, and 0.11$\degr$ respectively), so there is no shortage of good alignments to choose from.  With this, the dominant error will simply be the measurement error of $\sigma_{3-in-line}=0.10\degr$.

We can put limits on $\sigma_{3-in-line}$ from other considerations.  First, for the traditional resolution of the human eye being 1 arc-minute, we have $\sigma_{3-in-line}>0.017\degr$.  (Human resolution is actually a complicated issue, critically involving the star and background brightness, see Schaefer 1991.)  Second, humans can easily spot angular deviations of equal to the lunar radius, so $\sigma_{3-in-line}<0.25\degr$.  Third, the four reported great circle arcs all cross within 10 arc-seconds, and this suggests that $\sigma_{3-in-line}\sim0.003\degr$, although lucky coincidence might account for part of this close agreement.  Fourth, if the uncertainty is too large, then it becomes rather improbable that the first cross point (where $\beta$ Cep to $\gamma$ Cas crosses with $\iota$ Cep to $\delta$ Cas) is within 10 arc-seconds of the second cross point (where $\beta$ Cas to $\lambda$ UMa crosses with $\iota$ Cep to $\delta$ Cas).  The improbability of such a close match is roughly 0.003$\degr$/$\sigma_{3-in-line}$.  At the 100-to-1 probability level, we can say $\sigma_{3-in-line}<0.3\degr$.  These four limits are not very constraining, yet they can still give us confidence that our empirical measure is reasonable, so we adopt $\sigma_{3-in-line}=0.10\degr$.

\subsection{The Explosion Site}

With these measurement error bars, we have used the chi-square method to find the position on the sky which minimizes the $\chi^2$.  This best fit position is J2000 00:25:09.36 +64:08:49.  This position is 0.021$\degr$ (1.26 arc-minutes) to the north-west of the RL04 geometric center of the SNR.  The chi-square of this best fit position is 56.9, while the number of degrees of freedom is 42-2=40 (for a reduced chi-square of 1.42).  Recalling that the $\sigma$ values for four of the individual observers was set so that their reduced chi-squares are unity, the fact that the reduced chi-square is 1.42 implies that there must be some sort of systematic error between the observers, and that this systematic uncertainty must be comparable to the measurement errors.  In Table 1, for each astrometric measure, we give the angular distance from the star to the best fit position ($\Theta_{BestFit}$), the chi-square contribution for that point ($[(\Theta - \Theta_{BestFit})/\sigma]^2$), and the angle between the star and the geometric center of the SNR ($\Theta_{GeoCenter}$).  Fortunately, we see no points with excessive $\chi^2$ contribution, which is to say that there are no outliers, and that our final position is not being dominated by some singular error.

The one-sigma error region (i.e., the region on the sky with $\chi^2<\chi^2_{min}+1$) is close to an ellipse in shape.  The semi-major axis is 35 arc-seconds (pointing 60$\degr$ east of north, roughly on a northeast to southwest direction), while the semi-minor axis is 28 arc-seconds.  With this accuracy, close to half an arc-minute, the ancient astrometry suddenly becomes relevant to the modern astrophysics problem.  The three-sigma error ellipse (i.e., the region on the sky with $\chi^2<\chi^2_{min}+9$) has the same orientation, but with a semi-major axis of 105 arc-seconds and a semi-minor axis of 85 arc-seconds.  The position of star G is outside the 3-$\sigma$ region, and with a $\chi^2=69.9$, star G is rejected at the 3.6-$\sigma$ confidence level, for the measurement uncertainties alone.

This chi-square calculation gives correct error bars for measurement errors, but does not include possible systematic errors.  We have already had two indications that systematic errors are comparable to the measurement errors (that the reduced $\chi^2$ is 1.42 and that Tycho and Gemma both made corrections that somehow made their first three stars in perfect agreement).  Systematic errors, by their very nature, are hard to detect, especially now with the lack of any raw data.  Nevertheless, we can get a reasonable idea of the size of the systematic errors by looking at subsets of observers and by varying the choices in the analysis above.  That is, as the input and choice are varied, the best fit position will move around, and if these systematic errors are large then the positions will move around substantially, while if the systematic errors are small then the positions will vary little.  This is like a version of the resampling technique in statistics.  For this, we have taken eight alternative choices with variously reasonable changes to the selection of observers, the use of their original reported distances, and the assigned $\sigma$ values:  (1) As a test to see whether the poor observations of Munoz and Gemma are skewing the best fit position, our first variation is to use only the 29 measures from the other five observers.  (2) Tycho has the best observations, and our second variation is to use Tycho's data alone, as this makes us independent of any systematics by the other observers.  (3) Both Tycho and Gemma reported their final positions with corrections such that three stars are in perfect agreement, potentially with these corrections introducing systematic errors, so our third variation is to replace these six distances with their originally reported distances.  (4) Dreyer (1923) found positions of four stars from Tycho's observing logbook (Green 2004), although these may be only part of his data and not fully-corrected (e.g., for parallax of the eye behind the first slit), so our fourth variation is to use these four distances from Tycho's logbook.  (5) Hagecius corrected two of his distances, apparently after discussion with Tycho concerning the details of his instrumentation, so our fifth variation is to use Hagecius' original distances.  (6) Our estimate for the accuracy of Reisacher's single distance measure is that he was average with respect to the other observers, so our sixth variation is to take $\sigma_{Reisacher}=0.023\degr$.  (7) The accuracy of the three-star alignment observations is not well constrained, and we have already pointed to a reason to think that the real accuracy is better than we have adopted, so for our seventh variation, we have set $\sigma_{3-in-line}=0.03\degr$.  (8) Tycho's accuracy is very impressive, yet maybe the uncertainty is not so good, so our eighth variation is to double his uncertainty by setting $\sigma_{Tycho}=0.046\degr$.  These eight variations are essentially trying to see how robust is our final position.  The positions from these variations fall roughly within the one-sigma error region, so it is clear that the systematic errors are comparable to the measurement errors.  Quantitatively, the RMS scatter of these positions is 22 arc-seconds in both the directions along the semi-major and semi-minor axes.  This is not a perfect measure of the systematic uncertainties, but it is reasonable and it is a good measure of the robustness of the final position to variations.

The final error bars will be from both measurement and systematic errors.  Effectively, our analysis shows that the systematic errors ($\sigma_{sys}$) are an average of two-thirds times the observed measurement errors ($\sigma$).  Along the two axes, we can add the uncertainties in quadrature.  Along the semi-major axis, the total one-sigma error will be $(35^2+22^2)^{0.5}=42$ arc-seconds.  Along the semi-major axis, the total one-sigma error will be $(28^2+22^2)^{0.5}=36$ arc-seconds.    The relative sizes and orientations are shown in Figure 1.  With small loss of accuracy, we can say that the final position from the sixteenth century astrometry has an uncertainty of 39 arc-seconds.  

The error bars for the chi-square method are exact for the case where the error distribution is Gaussian, so we will now test for this condition.  (In practice, the error distributions can be rather far from Gaussian and still the best fit will be near perfect and the resultant error bars will be very good.)  We can test for a Gaussian distribution of $(\Theta - \Theta_{BestFit})/(\sigma^2+\sigma^2_{sys})^{0.5}$.  Out of the 42 measures in Table 1, 69.0\% have values between -1 and +1, 95.5\% have values between -2 and +2, while 100\% have values between -3 and +3.  These closely match that expected for a Gaussian distribution.  Thus, our quoted error bars for the position of the original supernova explosion have good accuracy. 

Critically, star G is 104 arc-seconds from the astrometry position, being roughly halfway between the semi-major and semi-minor axes, which is 2.6-$\sigma$ away.  That is, the sixteenth century astrometry alone rejects Star G being at the site of the explosion at the 2.6-$\sigma$ confidence level.  This conclusion has already included measurement and systematic errors, and is completely independent of offsets in the supernova remnant and proper motions.  While the rejection is not past the traditional three-sigma threshold, the 2.6-$\sigma$ rejection remains as a strong argument that star G is not the ex-companion star of the white dwarf that exploded in 1572.

\section{Explosion Position From a SNR Expansion Model}

Simplistically, the position of the supernova explosion should be at the geometric center of the SNR.  But no SNR is perfectly round, or even symmetric in shape, so it is unclear how to measure the center.  Further, it is unclear how to define the positions around the edge, and the center might depend on the wavelength of light used for the image.  Prior workers have reported central positions in the radio and X-ray (Duin \& Strom 1975; Henbest 1980; Reynoso et al. 1997; Reynoso \& Goss 1999;  Hughes 2000; Katsuda et al. 2010), but they reported little details as to their derivation, they gave no error bars, and indeed it appears that their reported centers were made with only approximate care.  To derive an accurate geometric center (and its error circle), we used a systematic centroid method to measure the centers from 8 images, all at different photon energies, each with a quantified error bar, and then combined these together, where our final geometric center uses the scatter of all 8 individual centers to define a final error circle.  The position of star G can then be compared to our final geometric center.

The position of the geometric center is actually not what is relevant for knowing the explosion position of the supernova, because there will inevitably be some offset between the two positions.  To see this, we can imagine an idealized case where the SNR is expanding into the ISM with a density gradient, so that one side is running into a higher gas/dust density than the other side.  The expansion towards the high density side will be slowed, while the expansion away from it will be relatively fast.  The SNR shape as seen from the side will have its geometric center at the point halfway from the slow edge and its fast edge.  In this case, the geometric center will be offset significantly from the explosion center in the direction away from the dense side.  Schaefer \& Pagnotta (2012) reported infrared bright edge being seen in SNR 0509-67.5, caused by differing amounts of swept up dust and mass in different directions, and this made for a flattened oval.  By fitting an ellipse, they determined the offset from the real SNe explosion site.  In general, relatively dense clouds around the edges will limit the expansion along those edges, making for complicated offsets.  Kaplan et al.  (2008) reported the explosion location of the Crab SNR through fitting a "divergent point" of the proper motion of many filaments in the remnant field.  In the case of Tycho's SNR, the remnant edges are roughly circular with relatively large deviations, while the swept up matter ( as judged by the brightness around the edge) is highly asymmetric.  The southern quarter has little swept-up mass, while relatively dense clouds are towards the northeast and towards the northwest.  With this, the 1572 explosion site must have an offset towards the north from the geometric center.  

A realistic model for the SNR should match the observed radii and expansion velocities in all directions from the center as a function of the mass of swept-up matter.  Fortunately, around the rim, Tycho's SNR has well measured expansion velocities (Katsuda et al.  2010 in X-ray; Reynoso et al.  1997 in radio), well measured densities for the swept-up material (Williams et al.  2013), and a realistic one-dimensional expansion model (Dwarkadas \& Chevalier 1998; Carlton et al.  2011), while our work on the geometric center has measured the radii around the rim from the geometric center which can be easily converted into radii from the explosion center.  With this, we can compare the observed and modeled radii and velocities as a function of the explosion position on the sky.  The best position will be the position with the lowest chi-square for this comparison, while the 1-$\sigma$ error region will be for places on the sky with a chi-square within 1.0 of this minimum.

\subsection{Geometric Center}

To determine the geometric center of the remnant, we have used public domain images in X-rays, infrared, and radio wavelengths.  Tycho's SNR is faint and ill-defined in optical light.  In X-rays, the {\it Chandra X-ray Observatory} provided images from 0.95-1.26 keV, 1.63-2.26 keV, and 4.1-6.1 keV.  In the far infrared, the {\it Wide-field Infrared Survey Explorer} (WISE) provided a 22 micron image, while the {\it Herschel Space Observatory} provided a 70 micron image.  In the radio, the Cambridge One-mile Telescope Radio provided images at 1.4 GHz in 1980, 2.7 GHz in 1972, and 5 GHz in 1972 (Henbest 1980, Tan \& Gull 1985).  The X-ray, radio, and infrared emission is all from different physical mechanisms within the remnant.  The reason for using all eight independent images is partly to beat down the error, but is mainly so that the our geometric center will not be dependent on the vagaries of the remnant's shape due to varying brightnesses changing with the emission mechanism.

In each image, we took the edge of the SNR to be the radial position at which the flux had fallen to 25\% of the peak flux in the nearby rim.  For each image, we started by constructing a set of nine baselines, with each baseline passing through a preliminary center and each with position angles at 10$\degr$ intervals.  Along each baseline, we determined the two edges of the SNR, bisected the resultant line segment, drew a perpendicular line through this middle point, determined the edge positions along this perpendicular line, and took the indicated center to be the middle position of the line segment on the perpendicular line.  This results in nine positions sampling different parts of the edge of the SNR.  We combined these nine points as a simple average to get the geometric center for this image.  (We iterated with this center in place of our preliminary center, but this made no difference in the resultant center.)  For a different analysis method, we fit circles to our 36 edge positions with a $\chi^2$ fit procedure, and these are in perfect agreement with the results we got from our first method.  

The uncertainty in the geometric centers for each image (from the nine baselines) is simply the RMS of the nine coordinates in both right ascension and declination, divided by the square root of 9 (for the number of independent centers averaged together).  All of the centers derived for the nine baselines produce positions consistent with a scatter of around 6", and so the resultant geometric center for each image has an uncertainty of around 2".  This is a smaller uncertainty than might be expected, but it is based on 36 measured edge positions for each image, and the small scatter in the nine independent centers for each baseline demonstrates that the geometric center can be determined to high accuracy for our given definition.

The eight images result in eight geometric centers (see Table 2 and Figure 2).  Our final geometric center is simply the straight average of all eight centers. (A weighted average gives the same final center and its error bar to within 0.3".)  The RMS scatter of these eight positions (2.5 arc-seconds in right ascension and 2.6 arc-seconds in declination) is dominated by the measurement errors.  Again, we see that the observed scatter in our 8 independent centers is smaller than might be expected, demonstrating that the geometric center has been measured to good accuracy for our given definition.  With this, the uncertainties in each coordinate are just the RMS scatter divided by the square-root of 8.  So our final geometric center is J2000 00h 25m 19.23s $\pm$ 0.12s and +64$\degr$ 08' 14.4 "$\pm$ 1.2".  

The uncertainty in our geometric center is greatly smaller than the search region in RL04.  Tycho star G is far to the east and the south of the geometric center (see Figure 2), being far outside the three-sigma error circle.  The proper motion of star G (Bedin et al. 2014) from 1572 to now is 2.1 arc-seconds, and is completely negligible in this context.

The geometric center is offset by some amount from the real explosion site of 1572.  The direction of this offset can be seen by looking at infrared images, with the brightness around the SNR rim being dominated by swept-up ISM dust, so the infrared bright portions of the rim indicate a lot of swept-up ISM material and a small expansion radius.  Tycho's SNR is infrared bright to the northeast and to the northwest, indicating that the 1572 explosion site must be somewhere north of the geometric center.  Similarly, the entire southern quadrant of the SNR is infrared faint, indicating little swept-up material, so that the expansion is relatively free and fast in that direction, again pointing to the true expansion center being somewhere to the north of the geometric center.  With star G being to the southeast of the our small geometric center error circle and with the explosion site being shifted in some amount to the north, there is no way that star G can be inside any error region for the explosion site.  Thus, the answer is already apparent from just our geometric center analysis plus the infrared image, and this is that star G cannot have been anywhere near the explosion site in 1572.

\subsection{Expansion Model}

To create a chi-square model for the radii and velocities around the edge of the SNR, we need a realistic physical model of the radius of the leading edge of the ejecta as a function of time and the density of swept-up material.  Fortunately, we can adopt an analytic expansion model, as used by other workers on Tycho's SNR (Dwarkadas \& Chevalier 1998; Carlton et al.  2011).  This is a one-dimensional model that assumes a constant density of swept-up ISM.  The model assumes that the SN ejecta has an exponential density profile given by 
\begin{equation}
\rho \propto t^{-3} exp\left(-v/v_e \right),
\end{equation}
where $t$ is the time since the start of the explosion, $v$ is the expansion velocity, and $v_e$ is the velocity scale for the supernova ejecta.  Carlton et al.  (2011) justified the applicability of thin shell approximation in a young supernova remnant, such as Tycho's SNR with age $\sim$450yrs.  They reached an analytic solution for the dimensionless blast wave radius as
\begin{equation} 
r^{'} \equiv \frac{R}{R^{'}}  = \left(1+4\frac{v_{e}}{v}+6\left(\frac{v_{e}}{v} \right)^2 \right)^\frac{1}{3} \left(\frac{2v}{v_{e}} \right)^\frac{1}{3} exp\left(\frac{-v}{3v_{e}} \right),
\end{equation}
\begin{equation}
R'=2.19 \left(\frac{M_e}{M_{ch}} \right)^{\frac{1}{3}} n_{0}^{-\frac{1}{3}} ~pc 
\end{equation}
\begin{equation} 
v/v_e = 8.03 (\frac{R}{2pc}) \left(\frac{t}{100yr} \right)^{-1} \left(E_{51} \right)^{-\frac{1}{2}} \left(\frac{M_{e}}{M_{ch}} \right)^\frac{1}{2}.
\end{equation}
Here, $E_{51}$ is the ejecta kinetic energy (in units of $10^{51}$ erg), $M_{e}/M_{ch}$ is the ejecta mass in units of the Chandrasekhar mass $M_{ch}$, and $n_{0}$ is the pre-shock ISM density in $cm^{-3}$.  Then, with given values of $t$, $E_{51}$, $M_e$, and $n_0$, we can solve for the radius $R$ and velocity $v$.

With this expansion model, we adopt $M_{e}=M_{ch}$, which is true for almost all models, and is approximately true even for the sub-Chandrasekhar models.  For the age of the SNR, we take $t=431$ years, as appropriate for the date of the measured expansion velocities.  Our measured radii are all revised to the year 2003, which correspond to the SNR age of 431 yr.  Based on prior published models, we expect that $E_{51}$ is somewhere between 0.3 and 1.75, with Hughes (2000) pointing to 0.4-0.5, but we treat this as a free parameter.

The swept-up material has a density that varies substantially around the edge of Tycho's SNR.  Badenes et al.  (2006) and Hayato et al.  (2010) found that ejecta towards the northeast and towards the northwest are brighter than those towards the south in X-rays, while similar brightenings are seen in the radio and infrared.  The directions towards the northwest and the northeast show a factor of 2-to-3 slower expansion rate compared to other directions (Fig.  5 in Katsuda et al.  2010; Fig.  4 in Tan \& Gull 1985; Fig.  6 in Reynoso et al.  1997).  Observation evidence show that Tycho's SNR might be interacting with dense molecular cloud on the eastern and northeastern sides, which gives bright emission and low expansion velocity (Decourchelle at al.  2001; Reynoso \& Goss 1999; Lee et al.  2004).  By tracing the outer edge of the cloud from photoionizing radiation, Ghavamian et al.  (2000) found that Tycho might be interacting with a warm ISM cloud.  Simulations by Chiotellis et al.  (2013) show models with a pre-existed stellar wind bubble would not only solve the discrepancy in ISM estimate but also match very well with the morphology, dynamics and x-ray spectrum of the current Tycho's SNR.  However they have trouble explaining the origin of the wind bubble.  It's pretty clear that Tycho's SNR shows evidence of asymmetric features that are apparently caused by the interaction with ambient ISM.  Williams et al.  (2013) performed a model fit for the observed 70 $\mu$m to 24 $\mu$m flux ratio from {\it Spitzer} to get the post-shock ISM density ($n_{post}$) distribution for 19 positions around the edge of the shell.  We adopted their values.  

Williams et al.  (2013) gives the post-shock density of the swept-up ISM, whereas $n_0$ (in equation 3) is the pre-shock density.  With the usual shock jump conditions for an adiabatic index of 5/3, the density will increase by a factor of 4 across the shock, so $n_{post}/n_0=4$.  For Tycho's SNR, the value might be somewhat different, for example from effects of cosmic ray acceleration and Rayleigh-Taylor instabilities (Vink et al.  2010; Williams et al.  2011;  Warren et al. 2005; Warren \& Blondin 2013).  Efficient cosmic ray acceleration would increase the shock compression ratio and bring the contact discontinuity closer to the forward shock.  Rayleigh-Taylor instabilities tend to smooth out high densities at the contact discontinuity that would affect the estimate of ISM density as well.  Both mechanisms have little or no impact on the expansion parameter (Warren \& Blondin 2013).  Since we do not know the exact shock jump density ratio, we take $n_{post}/n_0=Q$, where $Q$ is a free parameter.

\subsection{Chi-square Model Fit}

We now have a realistic expansion model for Tycho's SNR, where the radius and velocity can be predicted with adjustable parameters of $E_{51}$ and $Q$.  (Recall that we have set $t$=431 years, $M_e$=1, and take $n_0$ from Williams et al.  2013).  To convert the radii and velocities into angular radii and angular velocities, we further need the parameter of the distance to the SNR, $D$.  (This distance is a number that we will fit for, and we expect it to come out near the usual value of 2.3 kpc or so.)  Our expansion model must also specify the exact position of the supernova explosion, which we will label as $\alpha_{SN}$ and $\delta_{SN}$ for the right ascension and declination in J2000 coordinates.  Thus, our expansion model has five free parameters.

Our model can be compared to the observed radii and velocities.  The velocities for each position angle around the rim of the SNR are taken from Williams et al.  (2013), as shown in Table 3.  The radii at each position around the edge depends on the center, $\alpha_{SN}$ and $\delta_{SN}$.  For this, we have taken the positions of the edge (as in Section 3.1) and calculated their angular distances from the center for comparison with the model radii.  For our final best fit center, the observed radii of the SNR edge from this center for each position angle are also presented in Table 3.

We now have a complete expansion model that can be compared to a complete set of measured radii and velocities at 19 positions around the edge of Tycho's SNR.  For each set of five input model parameters, we can compare the model with the observations in the usual chi-square manner.  We can then vary around the five model parameters until we reach a minimum chi-square ($\chi^2_{min}$).  This set of five best parameters will then be our best fit, including our best fit position for the explosion site in 1572.  The 1-$\sigma$ uncertainty will be the ranges of the parameters for which $\chi^2 \leq \chi^2_{min}+1$, while the 3-$\sigma$ uncertainty will be the ranges of the parameters for which $\chi^2 \leq \chi^2_{min}+9$.

For our chi-square fit, we are comparing the observations with out model for 19 radii and 19 velocities.  Our model has 5 adjustable parameters; $E_{51}$, $Q$, $D$, $\alpha_{SN}$, and $\delta_{SN}$.  So our number of degrees of freedom is 33.  

Our measurement error for the radii are just a few arc-seconds, and in all models this leads to a large chi-square.  What is going on is that there is some systematic error that must be added in quadrature, and we think that this is due to natural variations in the SNR radii that cannot be now modeled.  In looking at the {\it Chandra} images, we see a very mottled surface with bumps and knobs on small and middling angular size scales, and these will make for intrinsic bumps as are seen around the edges.  Such bumpiness might arise from early instabilities in the ejecta outflow, and they are at smaller scales than we can get far infrared measures of the swept-up gas density for modeling.  To avoid problems arising from a wrongly-large chi-square and to recognize the reality that there are unmodeled natural systematic variations in the SNR profile, we have added in quadrature a systematic uncertainty.  The exact size of this added error is unknown, but we have set it such that the reduced chi-square associated with the measured radii is near unity.  Thus, our measured errors in the radii (as reported in Table 3) have been added in quadrature with 23 arc-seconds for use in our chi-square fit.  With this, the quality of our fit to the radii cannot be judged by our $\chi^2_{min}$.  Our procedure will not change the best fit parameters, but it will change the size of the error contours, making them reasonable for the reality of the bumps in the profile.

Our best fit has $\chi^2_{min}$=35.5.  This is for a distance $D$=2.30$\pm$0.04 kpc, $E_{51}$=0.46$\pm$0.04, $Q$=1.5$\pm$0.1; all reasonable values.  Our derived center is at J2000 coordinates of $\alpha_{SN}$=0h 25m 15.58s,  $\delta_{SN}$=64$^\circ$ 8'  39.8".  This newly determined supernova explosion site is to the 34.9 arc-seconds northwest of the geometric center.  The 1-$\sigma$ error region has a radius of 7.5 arc-seconds.  The 3-$\sigma$ error region has a 22.5 arc-second radius.  The relative sizes and positions of our expansion-model error regions are shown in Figure 1.  This result agrees with the estimated 40" shift from the geometric center of the remnant to the explosion site reported in Williams et al.  (2013).

The basic idea of our expansion model is to account for the variations in SNR radius, expansion velocity, and density of swept-up material in all the directions around the edge of the remnant.  All these quantities vary substantially with the position angle around the remnant.  Figure 3 and Figure 4 plot the radius and the expansion velocity for both the observed values as well as for the best fit model values.  The point of these figures is that the observed variations are reasonably well modeled, which is to say that our model is indeed catching the essence of the variations.  We see that the reason for the observed variations is largely due to the variations in the amounts of swept-up ISM material in different directions.

The ISM density around Tycho's SNR is not perfectly uniform.  Williams et al.  (2013) found that the northeast and northwest sides have over ten times higher densities than those to the south, which might indicating a density gradient in the pre-supernova ISM.  Any such overall gradient would require that any parcel of SN ejecta sweep up ISM material with density changes over the centuries since 1572.  The origin of the density gradient is unknown.  Chiotellis et al.  (2013) found the best scenario of the ambient ISM distribution to reproduce the observed morphology, dynamics and X-ray emission spectrum of Tycho's SNR, which is that the supernova shock evolved into a pre-existed stellar wind bubble but is now expanding into a uniform ISM with lower density.  This would reconcile the discrepancy between two ISM density determinations, where the current high shell expansion velocity and lack of thermal X-ray emission in the shell put a upper limit on the ISM density of $\sim$0.6 cm$^{-3}$ (Cassam-Chena et al.  2007; Katsuda et al.  2010), while the high ionization age of Tycho's SNR derived from X-ray spectrum requires a high ambient ISM density.  The resulting density of the wind and homogenous ISM is well below the upper limit.  However the origin of the wind bubble remains a mystery.  Stellar wind from low or intermediate mass giant stars suffers from lacking the detection of any giant stars within the centroid of Tycho's SNR that has the required properties.  Also, the derived duration for the mass outflow is $\sim$5$\times$10$^{4}$ years, and this is greatly shorter than the life time of giant stars, so fine tuning would be needed.  Recurrent nova could generate a sequence of explosion that eject its shells and the collision among them might produce the density structure seen in their model.  On the other hand, for a DD channel, whether the ejection of common envelope  could recreate the same feature remains unclear (Chiotellis et al.  2013).  Figure 4 in Chiotellis et al.  (2013) shows that around $\sim$500 yr, the remnant evolves into a wind bubble that resembles the same features (morphology and dynamics) produced by one with a uniform ISM.  Nonuniform ISM near to the SN will have negligible effect on the model expansion results, because there is little material swept up when compared to the mass of the ejecta.  The effects of the progenitor (from wind bubbles, recurrent nova shells) will generally provide substantial density variations only close to the progenitor.

To compensate for these potential variations in the ISM density, we allow the parameter Q to vary freely, instead of fixing it to some specific value (such as a standard shock jump ratio) in our calculation.  Our best fit value of Q is below the standard value of 4, which might be telling us that Williams et al.  (2013) has underestimated the post-shock density or that Rayleigh-Taylor instabilities are important.  More fundamentally, our analysis is assuming that the relative densities around the shell are fixed and all the physical mechanisms affect the dynamics of the SNR in the same way in all directions.  We are fitting the remnant in a relative way, with the shift between centroid of the SNR and real explosion site being independent of this assumption.  

\subsection{Explosion Site From a Simple Model}

We can make a test for the sensitivity of our derived center position to our adopted expansion model.  For this, we have adopted a rather simple expansion model, in this case simply requiring the conservation of momentum of the ejecta, all in a thin shell, as it sweeps up ISM material.  As a simple numerical integral, we get the model shell radius and velocity as a function of the distance, the ejecta mass, the ejecta velocity, and the ISM density.  We substituted this simple model for the expansion model from equations 2-4, we calculated the model radius and shell velocity along the same 19 azimuthal angles, and we compared those with the observed radii and velocities as discussed in the previous section.  For this chi-square comparison, the position of the explosion site on the sky provides two further free parameters, and indeed are the two parameters of interest here.  By varying all the input parameters, we get a minimum $\chi^2$ with $\alpha_{SN}$=0h 25m 15.34s,  $\delta_{SN}$=64$^\circ$ 8' 40.9" which is just 2.1" away from the location we got from a more complex model.  We take the independence in the derived supernova centers with a greatly different expansion model to be a good argument that our method is robust on the details of the expansion model.

\subsection{Effects of an Inhomogeneous ISM}

For the previous calculations, in both the simple and the realistic models, we have assumed that the density of the swept-up ISM to be a constant in the expansion history.  While this is reasonable as an approximation, it is fully possible that any fragment of the expanding shell can encounter a significantly inhomogeneous ISM, for example as the supernova might blow up inside some sort of a bubble from the progenitor or the ejecta might encounter a relatively dense cloud in some direction.  So we should calculate the sensitivity of the derived explosion site to plausible inhomogeneous ISM distributions.  For this, we make the calculations within both the realistic model (cf. Section 3.2) and the simple model (c.f. Section 3.4).

Considering the ISM in each direction to have a density that is a step function, where its density is $n_I$ out to some inner radius $R_{I}$, while outside that radius it has a density of $n_{0}$.  The total swept up ISM mass will be $(4\pi/3)(n_{0}(R_{s}^{3}-R_{I}^{3})+n_IR_{I}^{3})$, where $R_{s}$ is the shell radius.  By introducing $\eta$ as the ratio between $n_I$ and $n_0$ and $\gamma$ as the ratio between $R_I$ and $R_s$, the swept-up ISM mass can now be written as  $(4\pi/3)(1-\gamma^3+\eta\gamma^3)n_{0}R_{s}^{3}$.  This simple model is not the same as complex situations that can be envisioned (e.g., see Chiotellis et al. 2013), but nevertheless, a wide range in choices of $\gamma$ and $\eta$ can demonstrate the level of sensitivity in the derived explosion site to the inhomogeneities of the ISM.  Following a similar procedure and notations described in Carlton et al. (2011), we can reach an analytic, parametric solution for the SNR shell radius.  This analytic solution has 
\begin{equation}
R'=2.19(1-\gamma^3+\eta\gamma^3) \left(\frac{M_e}{M_{ch}} \right)^{\frac{1}{3}} n_{0}^{-\frac{1}{3}} ~pc,
\end{equation}
along with equations 2 and 4.  Note that when $\gamma=0$ or $\eta=1$, the solution will reduce to that of the homogenous ISM density profile.  

With this, we can then repeat our $\chi^2$ calculations of the position of the explosion site for a variety of ISM distributions.  When $\gamma$=0.5, $\eta$=0, which means the SN was expanding into a bubble, we get a minimum $\chi^2$ with $\alpha_{SN}$=0h 25m 15.10s,  $\delta_{SN}$=64$^\circ$ 8' 42.36".  For $\gamma$=0.5, $\eta$=0.5, we get a minimum $\chi^2$ with $\alpha_{SN}$=0h 25m 15.20s,  $\delta_{SN}$=64$^\circ$ 8' 42.00".  For $\gamma$=0.5, $\eta$=2, the SN ejecta was sweeping through a dense ISM then encountered less dense material, we get a minimum $\chi^2$ with $\alpha_{SN}$=0h 25m 15.60s,  $\delta_{SN}$=64$^\circ$ 8' 38.76".  In all these cases, the reduced $\chi^2$ is close to unity.  All these positions are within a few arc-seconds from our final position (Section 3.3) and are within the one-sigma error circle of 7.5 arc-seconds.  However, for an extreme case with $\gamma$=0.8, $\eta$=0, which resembles the idea that Tycho's SN ejecta only went into some ISM very recently, we get $\alpha_{SN}$=0h 25m 14.35s,  $\delta_{SN}$=64$^\circ$ 8' 48.12". This is 10 arc-seconds to the north-west of the position from Section 3.3, which is even further away from Star G.  But in this case, we are getting a much larger minimum $\chi^2$.  All this is saying that the position of the derived center has a small sensitivity on the radial distribution of the ISM material being swept-up, where the change of position is comparable and less than the one-sigma uncertainty.  This conclusion is similar to the result in Figure 4 of Chiotellis et al. (2013), where the complex wind-driven bubble makes for little change from a homogenous ISM case in terms of the observed shell radius.

We have also examined the case of ISM inhomogeneities within the simple model of section 3.4. By adding different ISM density profiles, for example, a bubble or changing density along radius, we are still getting expansion centers within a few arc-seconds of the site given in section 3.4.   

We conclude that substantial density inhomogeneities along the radii of expansion lead to shifts in the derived explosion site that are insensitive to the expansion model.  The results from both models shows that the derived center is insensitive to even substantial inhomogeneities in the ISM.  With this, we adopt the best position from expansion models as that from section 3.3.

\subsection{The Ex-companion Candidates}

For the position of Tycho star G in 1572, the best chi-square is 102.9, with D=2.33$\pm$0.04 kpc, $E_{51}$=0.45$\pm$0.03, Q=1.55$\pm$0.14.  The position of star G has $\chi^2-\chi^2_{min}$=67.4=8.2$^2$.  Thus star G is rejected at the 8.2-$\sigma$ confidence level.  Similarly, ex-companion candidate star B is rejected at the 5.1-$\sigma$ level, and star E is rejected at the 4.1-$\sigma$ level.  See Figure 1 for the relative placement of these stars and our two independent error regions.

\section{Conclusions}

We have answered the question of the position of the 1572 supernova, and we have answered it with two positions.  Both positions are significantly offset from the geometric center.  The good agreement between our two positions, with radically different input, provides substantial confidence that {\it both} methods are accurate to within the stated error bars.  That is, it is very unlikely that both methods would suffer significant unknown systematic errors that moved both positions by a similar amount in a similar direction.  

With two valid methods to measure the position of the explosion site in 1572, our best measure will be the weighted average of the two positions.  With this, our final position is $\alpha$=0h 25m 15.36s,  $\delta$=64$^\circ$ 8' 40.2", and the 1-$\sigma$ error radius is 7.3 arc-seconds.  Star G is rejected at the 8.2-$\sigma$ level.  This proves a final and confident resolution for the controversy of whether this star is the ex-companion.  Further, star B is rejected at the 5.1-$\sigma$ level, while star E is rejected at the 4.1-$\sigma$ level.

We now have a confident and small error circle (see Figure 5).  This is the region for which any ex-companion must be sought.  Unfortunately, prior work has been exclusively inside the RL04 error circle, shown in Figure 2.  We see that a large part of the new error circle has not been examined for ex-companions.  That is, all prior searches have largely been looking in the wrong place.  A new search is required to answer the question whether Tycho's SNR has any ex-companion star.

Prior work has provided photometry of some stars inside our final error circle (Bedin et al.  2014).  These are labeled stars N, O, P, Q, R, and S (see Figure 3).  Both stars O and P are given as being early-G spectral type, and both were rejected as ex-companions because their distances would be much closer than the SNR if they are main sequence stars, while their distances would be much farther than the SNR if they are typical red giants (Bedin et al.  2014).  But this analysis ignores the possibility that either star O or P is a sub-giant, in which case their distance would match that of the SNR.  (This is the exact same idea used to put star G at the distance to the SNR.  Indeed, stars G, O, and P have identical colors and similar magnitudes, so any effort to put star G inside the remnant must also work for stars O and P.)  We note that Bedin et al.  (2014) measured star O to have a proper motion only a bit smaller than star G, so star O appears to be about as good an ex-companion candidate as star G originally was.  And who knows what other candidates lie within our new error circle?

~

This work is supported under a grant from the National Science Foundation.

{}

\begin{deluxetable}{lllllll}
\tabletypesize{\scriptsize}
\tablewidth{0pc}
\tablecaption{Astrometry of Tycho's Supernova}
\tablehead{\colhead{Observer} & \colhead{Star(s)} & \colhead{$\Theta$} & \colhead{$\sigma$} & \colhead{$\Theta_{BestFit}$} & \colhead{$\chi^2$} & \colhead{$\Theta_{GeoCenter}$}}
\startdata
Tycho	&	$\alpha$ Cas	&	7$\degr$ 50.5'	&	0.023	&	7.83	&	0.09	&	7.82	\\
Tycho	&	$\beta$ Cas	&	5$\degr$ 19'	&	0.023	&	5.34	&	1.30	&	5.34	\\
Tycho	&	$\gamma$ Cas	&	5$\degr$ 2'	&	0.023	&	5.00	&	2.02	&	4.98	\\
Tycho	&	$\delta$ Cas	&	8$\degr$ 3.5'	&	0.023	&	8.02	&	2.16	&	8.00	\\
Tycho	&	$\epsilon$ Cas	&	9$\degr$ 48'	&	0.023	&	9.77	&	1.81	&	9.75	\\
Tycho	&	$\zeta$ Cas	&	10$\degr$ 22'	&	0.023	&	10.36	&	0.14	&	10.35	\\
Tycho	&	$\eta$ Cas	&	6$\degr$ 53'	&	0.023	&	6.85	&	1.89	&	6.84	\\
Tycho	&	$\iota$ Cas	&	12$\degr$ 58.5'	&	0.023	&	12.97	&	0.02	&	12.96	\\
Tycho	&	$\kappa$ Cas	&	1$\degr$ 31'	&	0.023	&	1.50	&	0.80	&	1.48	\\
Tycho	&	$\alpha$ UMi	&	25$\degr$ 14'	&	0.023	&	25.23	&	0.04	&	25.24	\\
Tycho	&	$\alpha$ Per	&	27$\degr$ 22'	&	0.023	&	27.44	&	10.02	&	27.42	\\
Tycho	&	$\alpha$ Aur	&	42$\degr$ 28'	&	0.023	&	42.52	&	5.83	&	42.51	\\
Hagecius	&	$\alpha$ Cas	&	7$\degr$ 47'	&	0.153	&	7.83	&	0.11	&	7.82	\\
Hagecius	&	$\beta$ Cas	&	5$\degr$ 15'	&	0.153	&	5.34	&	0.37	&	5.34	\\
Hagecius	&	$\gamma$ Cas	&	5$\degr$ 3'	&	0.153	&	5.00	&	0.10	&	4.98	\\
Hagecius	&	$\eta$ Cas	&	7$\degr$ 0'	&	0.153	&	6.85	&	0.94	&	6.84	\\
Hagecius	&	$\kappa$ Cas	&	1$\degr$ 26'	&	0.153	&	1.50	&	0.17	&	1.48	\\
Hagecius	&	$\alpha$ UMi	&	25$\degr$ 30'	&	0.153	&	25.23	&	3.15	&	25.24	\\
Munoz	&	$\alpha$ Cas	&	7$\degr$ 50'	&	0.82	&	7.83	&	0.00	&	7.82	\\
Munoz	&	$\beta$ Cas	&	5$\degr$ 20'	&	0.82	&	5.34	&	0.00	&	5.34	\\
Munoz	&	$\gamma$ Cas	&	5$\degr$ 10'	&	0.82	&	5.00	&	0.04	&	4.98	\\
Munoz	&	$\alpha$ UMi	&	26$\degr$ 40'	&	0.82	&	25.23	&	3.08	&	25.24	\\
Gemma	&	$\alpha$ Cas	&	7$\degr$ 24'	&	0.38	&	7.83	&	1.31	&	7.82	\\
Gemma	&	$\beta$ Cas	&	5$\degr$ 4'	&	0.38	&	5.34	&	0.53	&	5.34	\\
Gemma	&	$\gamma$ Cas	&	4$\degr$ 36'	&	0.38	&	5.00	&	1.11	&	4.98	\\
Gemma	&	$\zeta$ Cas	&	9$\degr$ 36'	&	0.38	&	10.36	&	3.98	&	10.35	\\
Gemma	&	$\eta$ Cas	&	6$\degr$ 36'	&	0.38	&	6.85	&	0.44	&	6.84	\\
Gemma	&	$\kappa$ Cas	&	1$\degr$ 24'	&	0.38	&	1.50	&	0.06	&	1.48	\\
Gemma	&	$\alpha$ UMi	&	24$\degr$ 40'	&	0.38	&	25.23	&	2.19	&	25.24	\\
Gemma	&	$\alpha$ Per	&	27$\degr$ 7'	&	0.38	&	27.44	&	0.72	&	27.42	\\
Gemma	&	$\alpha$ Aur	&	42$\degr$ 4'	&	0.38	&	42.52	&	1.44	&	42.51	\\
Reisacher	&	$\kappa$ Cas	&	1$\degr$ 25'	&	0.136	&	1.50	&	0.34	&	1.48	\\
Digges	&	$\alpha$ Cas	&	7$\degr$ 47'	&	0.042	&	7.83	&	1.49	&	7.82	\\
Digges	&	$\beta$ Cas	&	5$\degr$ 15'	&	0.042	&	5.34	&	4.89	&	5.34	\\
Digges	&	$\gamma$ Cas	&	4$\degr$ 58'	&	0.042	&	5.00	&	0.65	&	4.98	\\
Digges	&	$\delta$ Cas	&	8$\degr$ 5'	&	0.042	&	8.02	&	1.96	&	8.00	\\
Digges	&	$\epsilon$ Cas	&	9$\degr$ 45'	&	0.042	&	9.77	&	0.21	&	9.75	\\
Digges	&	$\kappa$ Cas	&	1$\degr$ 28.5'	&	0.042	&	1.50	&	0.25	&	1.48	\\
Digges	&	$\beta$ Cep to $\gamma$ Cas	&	90$\degr$ \tablenotemark{a}	&	0.1	&	89.96	&	0.15	&	89.97	\\
Digges	&	$\iota$ Cep to $\delta$ Cas	&	90$\degr$ \tablenotemark{b}	&	0.1	&	89.99	&	0.02	&	89.99	\\
Maestlin	&	$\iota$ Cep to $\delta$ Cas	&	90$\degr$ \tablenotemark{b}	&	0.1	&	89.99	&	0.02	&	89.99	\\
Maestlin	&	$\beta$ Cas to $\lambda$ UMa	&	90$\degr$ \tablenotemark{c}	&	0.1	&	89.90	&	1.05	&	89.92	\\
\enddata
\tablenotetext{a}{90$\degr$ from J2000 22h 47m 15.6s -19$\degr$ 26" 17'}
\tablenotetext{b}{90$\degr$ from J2000 22h 47m 15.6s -23$\degr$ 47' 06"}
\tablenotetext{c}{90$\degr$ from J2000 16h 56m 41.3s +10$\degr$ 27' 04"}
\label{Table1}
\end{deluxetable}

\begin{deluxetable}{lll}
\tabletypesize{\scriptsize}
\tablecaption{Geometric center of Tycho's SNR
\label{tbl2}}
\tablewidth{0pt}
\tablehead{
\colhead{ID}                   &
\colhead{$\alpha$ (J2000)}                   &
\colhead{$\delta$ (J2000)}                   

}

\startdata
RL04 geometric center&	0h 25m 19.9s  & 64$\degr$ 08' 18.2"\\
WISE (22 $\mu$m)&	0h 25m 19.63s $\pm$ 0.33s&	64$\degr$ 08' 13.1" $\pm$ 2.6"\\
Herschel (70 $\mu$m)&	0h 25m 18.85s $\pm$ 0.45s&	64$\degr$ 08' 9.3" $\pm$ 1.1"\\
Chandra X-ray (950-1260 eV)&	0h 25m 19.35s $\pm$ 0.42s&	64$\degr$ 08' 13.3" $\pm$ 1.9"\\
Chandra X-ray (1630-2260 eV)&	0h 25m 19.42s $\pm$ 0.41s&	64$\degr$ 08' 12.9" $\pm$ 2.2"\\
Chandra X-ray (4100-6100 eV)&	0h 25m 19.33s $\pm$ 0.39s&	64$\degr$ 08' 12.7" $\pm$ 2.3"\\
Radio (1.4 GHz)&	0h 25m 18.54s $\pm$ 0.29s&	64$\degr$ 08' 18.7" $\pm$ 2.2"\\
Radio (2.7 GHz)&	0h 25m 19.10s $\pm$ 0.36s&	64$\degr$ 08' 17.9" $\pm$ 2.6"\\
Radio (5 GHz)&	0h 25m 19.58s $\pm$ 0.32s& 64$^\circ$ 08' 17.5" $\pm$ 2.3"\\
Combined geometric center&	0h 25m 19.23s $\pm$ 0.13s&	64$\degr$ 08' 14.4" $\pm$ 1.2" \\
\enddata
\end{deluxetable}

\begin{deluxetable}{cccccc}
\tabletypesize{\scriptsize}
\tablecaption{Observed radii and velocities, and their best fit model values
\label{tbl3}}
\tablewidth{0pt}
\tablehead{
\colhead{$\Theta$\tablenotemark{a}}                   &
\colhead{$n_0$ (cm$^{-3}$)\tablenotemark{b}}                   &
\colhead{$R$ (arcsec)\tablenotemark{c}}                   &
\colhead{$v$ (arcsec/yr)\tablenotemark{d}}                   &
\colhead{Model $R$ (arcsec)\tablenotemark{e} }                   &
\colhead{Model $v$ (arcsec/yr)\tablenotemark{f}}                   
}

\startdata
13&	0.22&	227.2&	0.335&	246.2&	0.307\\
32&	0.25&	217.0&	0.303&	240.7&	0.297\\
47&  0.40&	223.9&	0.216&	224.3&	0.269\\
63&	0.82&	212.4&	0.176&	199.0&	0.226\\
81&	1.44&	209.9&	0.203&	180.8&	0.197\\
105&	0.29&	247.9&	0.285&	236.0&	0.289\\
121&	0.27&	260.5&	0.322&	237.8&	0.292\\
138&	0.21&	246.0&	0.305&	247.4&	0.309\\
155&	0.17&	251.8&	0.319&	255.5&	0.324\\
172&	0.17&	265.9&	0.297&	255.5&	0.324\\
192&	0.08&	285.7&	0.346&	284.4&	0.377\\
213&	0.08&	283.9&	0.372&	288.0&	0.383\\
233&	0.08&	291.9&	0.365&	288.0&	0.383\\
252&	0.08&	285.4&	0.359&	284.4&	0.377\\
272&	0.09&	276.2&	0.353&	281.2&	0.371\\
290&	0.13&	267.3&	0.339&	266.2&	0.343\\
308&	0.37&	252.9&	0.328&	226.9&	0.273\\
331&	1.10&	225.7&	0.293&	189.5&	0.211\\
353&	0.45&	236.1&	0.218&	220.2&	0.262\\

\enddata
\tablenotetext{a}{Position angle around the edge of the remnant, measured from north towards the east}
\tablenotetext{b}{Pre-shock ISM density, taken from the Williams et al. (2013) measured post-shock density divided by our best fit Q=1.5.}
\tablenotetext{c}{Radius measured from remnant edge to the best fit explosion position, see Figure 3}
\tablenotetext{d}{Remnant expansion velocity measured from X-ray and radio observations, from Williams et al. (2013), with our best fit distance of 2.3 kpc, see Figure 4}
\tablenotetext{e}{Radius from the best fit model, see Figure 3}
\tablenotetext{f}{Remnant expansion velocity from the best fit model, converted to units of arcsec/yr with our best fit distance of 2.3 kpc, see Figure 4}
\end{deluxetable}

\begin{figure}
	\centering
         \makebox[\textwidth][c]{\includegraphics[width=1.2\textwidth, height=110mm, keepaspectratio]{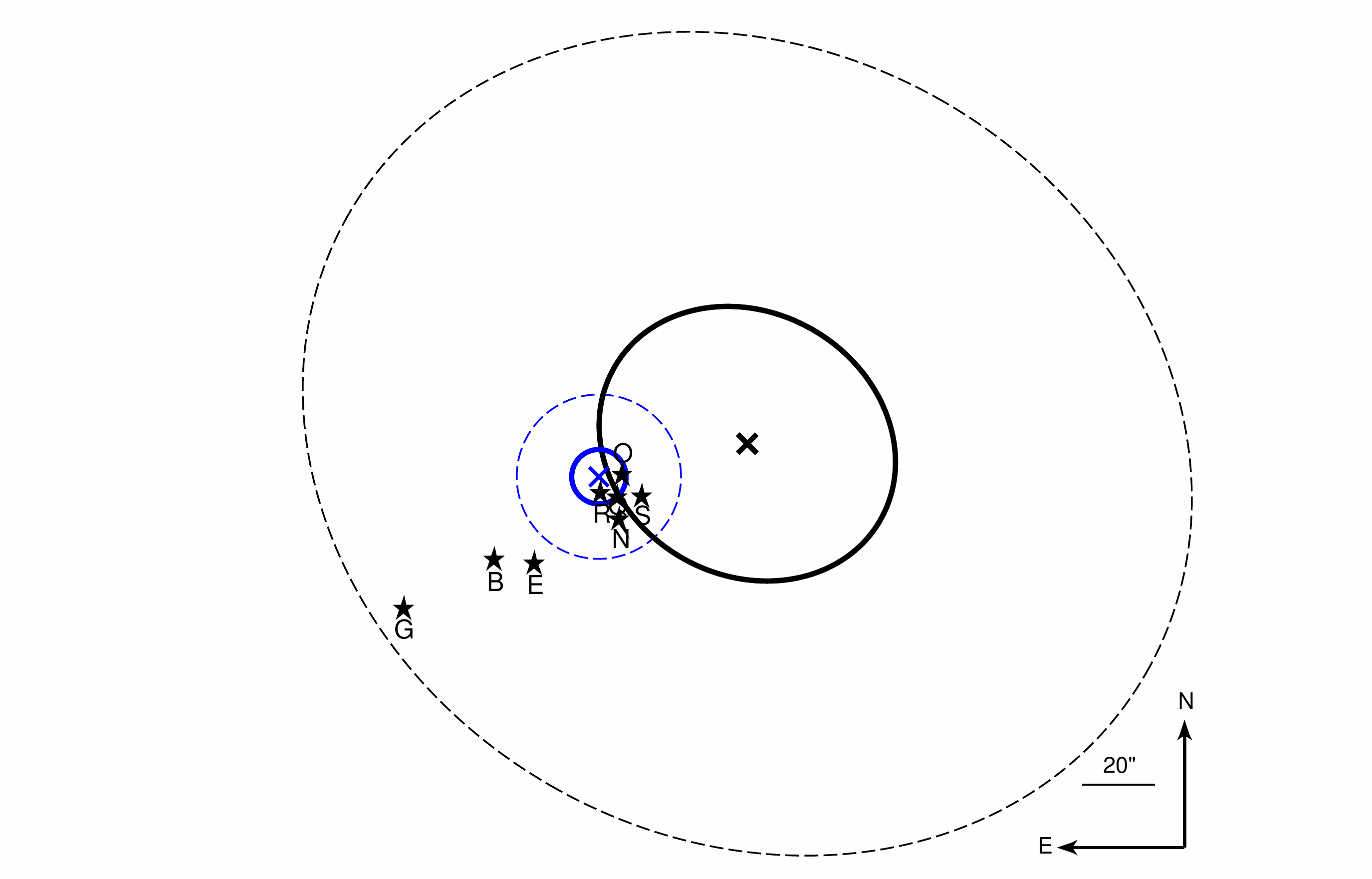}}	
         \caption{The two new and independent measured positions for the site of the 1572 supernova.  Our first method is to use the 42 astrometric measures of the supernova itself as observed by Tycho himself, plus six other observers, in 1572.  Our best fit position is shown with the black X, the thick tilted ellipse is our one-sigma error region, while the large thin-dashed ellipse is the three-sigma error region.  No prior workers had made a serious attempt to use the original astrometry to get the position of SN 1572, and it turns out to be surprisingly good.  Our second method is to use a realistic expansion model, as applied to 19 points on the outer edge of Tycho's SNR with measured radii, expansion velocities, and post-shock ISM densities.  Our best fit position for this second method is marked with a blue X, the thick-lined blue circle represents the one-sigma error region (with a 7.5 arc-second radius), and the thin-dashed blue circle encloses the three-sigma error region.  A variety of points can be seen from this Figure:  (1) The two greatly-different methods are in good agreement with each other, with their one-sigma regions having much overlap, and in good agreement that SN 1572 exploded to the NW of the modern SNR geometric center.  This provides good confidence that both methods are free from any substantial systematic error.  (2) Star G is rejected at the 2.6-sigma level and the 8.2-sigma level for the two methods.  This provides a simple and sure resolution to the long-standing controversy as to whether this star is the ex-companion.  (3) Star B and E, other proposed ex-companion candidates, are also rejected at the 5.1-sigma and 4.1-sigma levels, respectively.  (4) The real site of SN 1572 is a small region around stars O and R.}
\end{figure}

\begin{figure}
	\centering
	\makebox[\textwidth][c]{\includegraphics[width=1.2\textwidth]{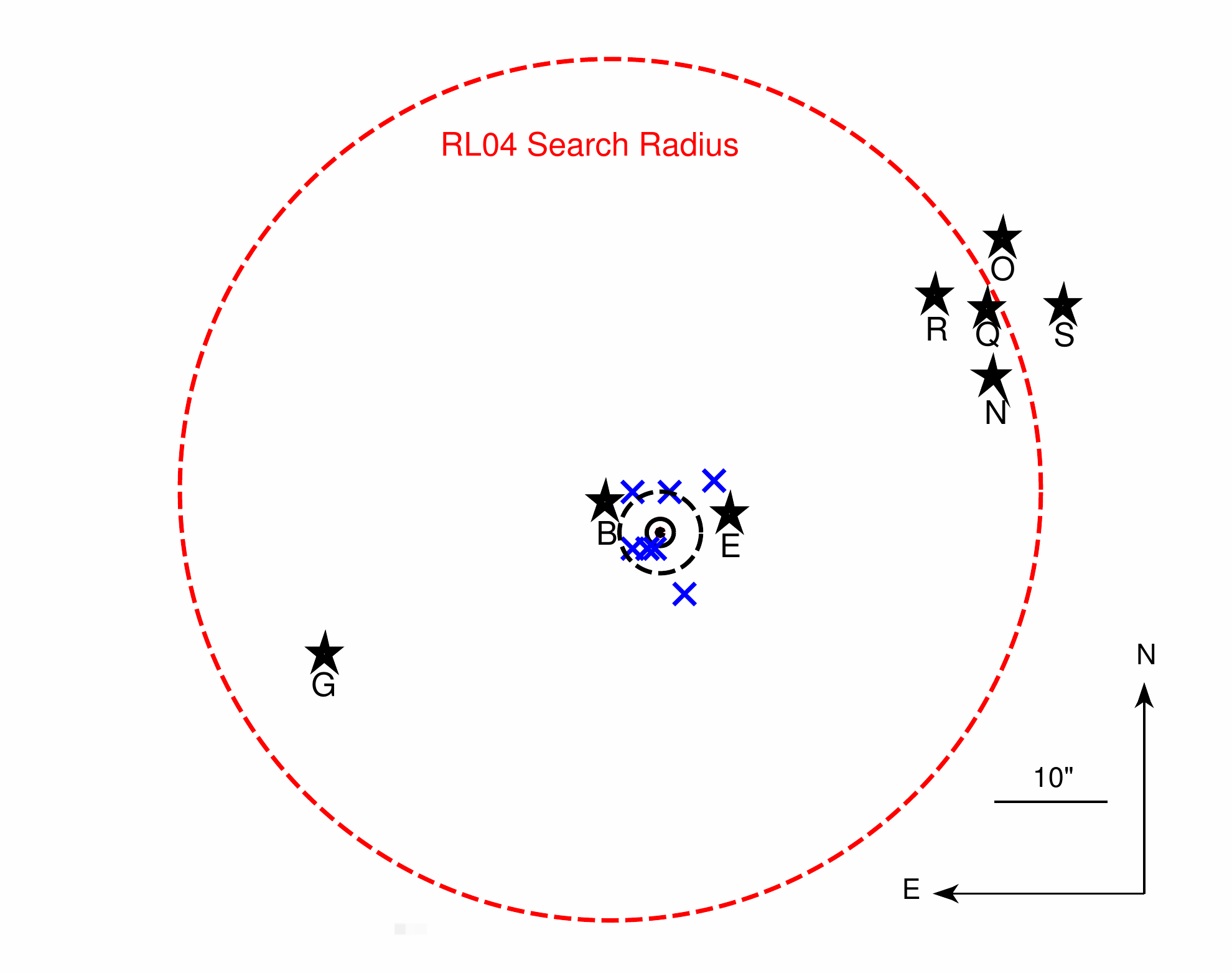}}
	\caption{Geometric center of Tycho's SNR.  The dashed red circle is the original search radius in RL04 for ex-companion candidates with a 0.65 arc-min radius.  Previously proposed ex-companion stars, Tycho stars G, B, and E, are within that original search region.  The eight blue Xs are the eight geometric center derived from infrared, X-ray and radio images.  The black dot is the combined geometric center, while the solid circle is the one-sigma error region for this geometric center (with 1.2 arc-second radius), and the black thin dashed circle shows the three-sigma error region.  We know from the infrared image that the SNR has little swept-up material towards the south, so the explosion site must be northwards from the geometric center.  But star G is to the south and east of the geometric center, so putting in the offset to get to the explosion site will certainly have star G far away.}
\end{figure}

\begin{figure}
	\centering
	\makebox[\textwidth][c]{\includegraphics[width=1.2\textwidth]{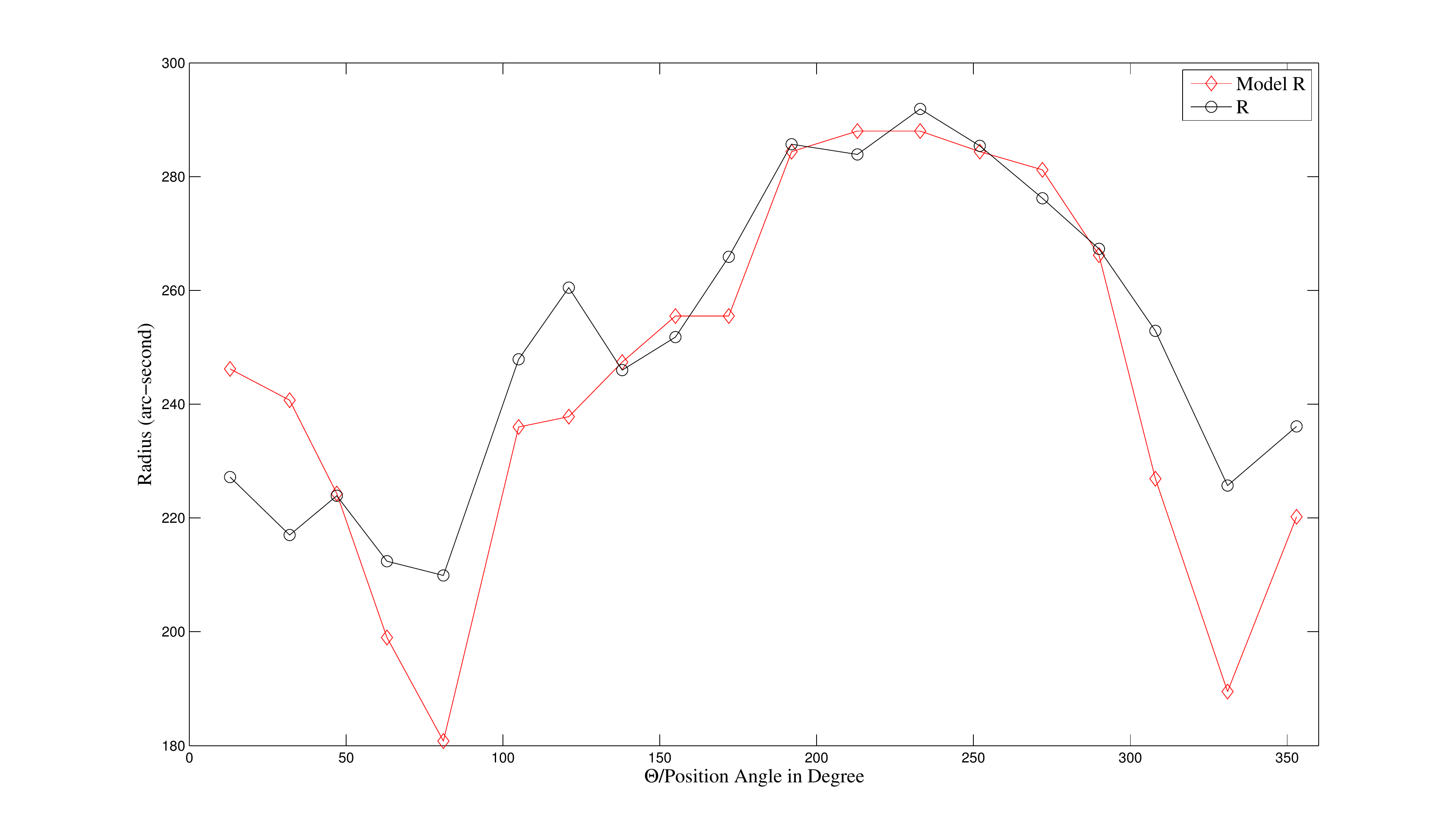}}
	\caption{The 19 observed radii and best fit model radii from SNR expansion model.  All values are measured or calculated with respect to the best fit explosion site.  The point of this figure is that the observed radii vary substantially around the edge of Tycho's SNR, and our modeled radii match these variations reasonably well.  We do not expect arc-second agreement, because the outer edge of the remnant is partly determined by turbulent features not perfectly represented or resolved by the measures of the density of the swept-up material.}
\end{figure}

\begin{figure}
	\centering
	\makebox[\textwidth][c]{\includegraphics[width=1.2\textwidth]{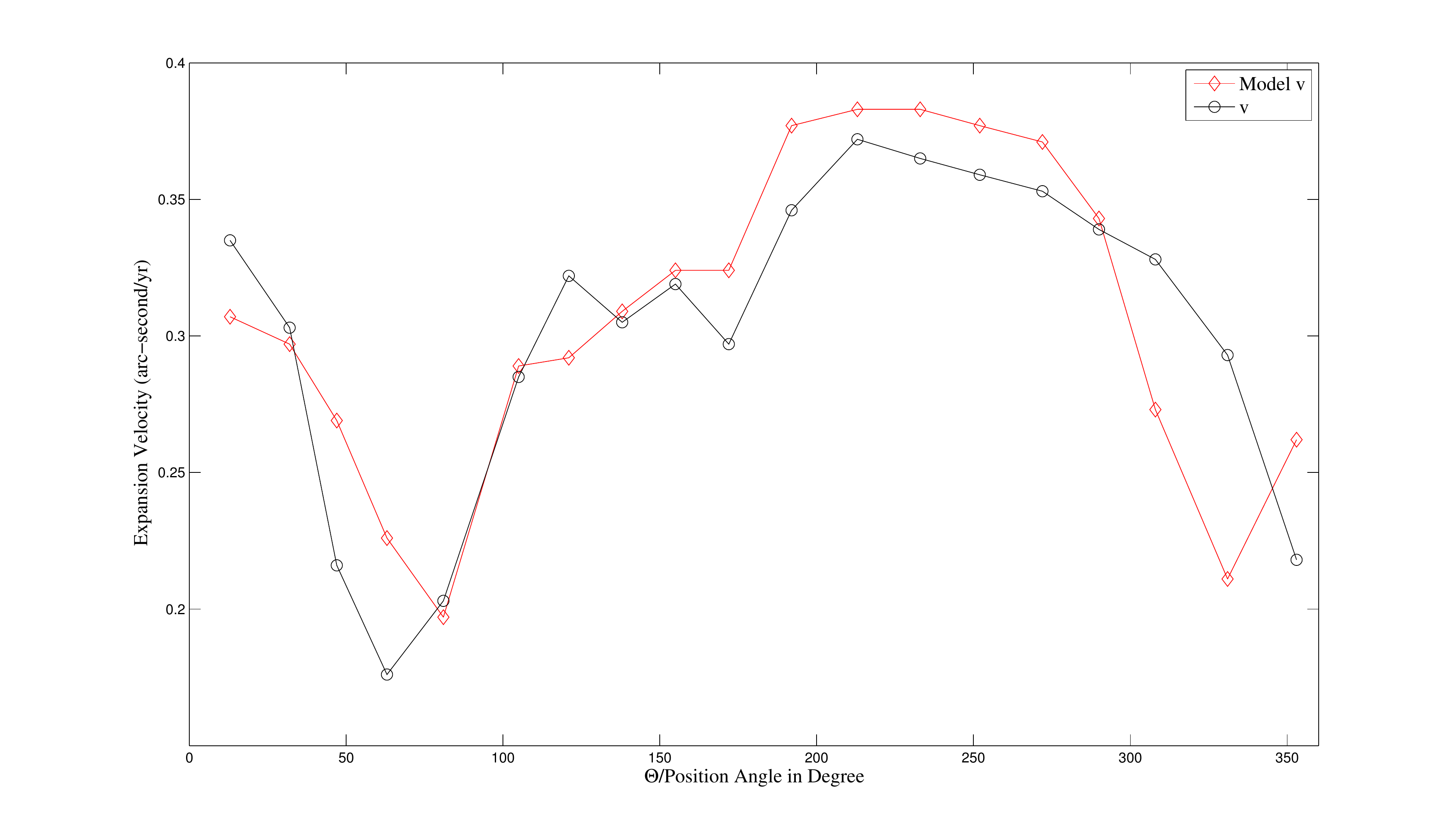}}
	\caption{The 19 observed SNR expansion velocity and the model expansion velocities.  Again, the expansion velacities vary greatly around the edges of the SNR, corresponding closely with the amount of swept up ISM material, and our model closely follows the observations.  In comparing Figures 3 and 4 with Table 3, we see that the SNR has low expansion velocities in the same directions that it has small radii, and these are the same directions with dense ISM material.  We understand the physics of this, with the detailed expansion models of Carlton et al. (2011) allowing us to work backwards in time to determine the original center of expansion.  Figures 3 and 4 are displays that our model is indeed matching the observed expansion history of the SNR, and hence that we can derive an accurate SNR expansion center.}
\end{figure}

\begin{figure}
	\centering
         \makebox[\textwidth][c]{\includegraphics[width=1.2\textwidth, height=122mm, keepaspectratio]{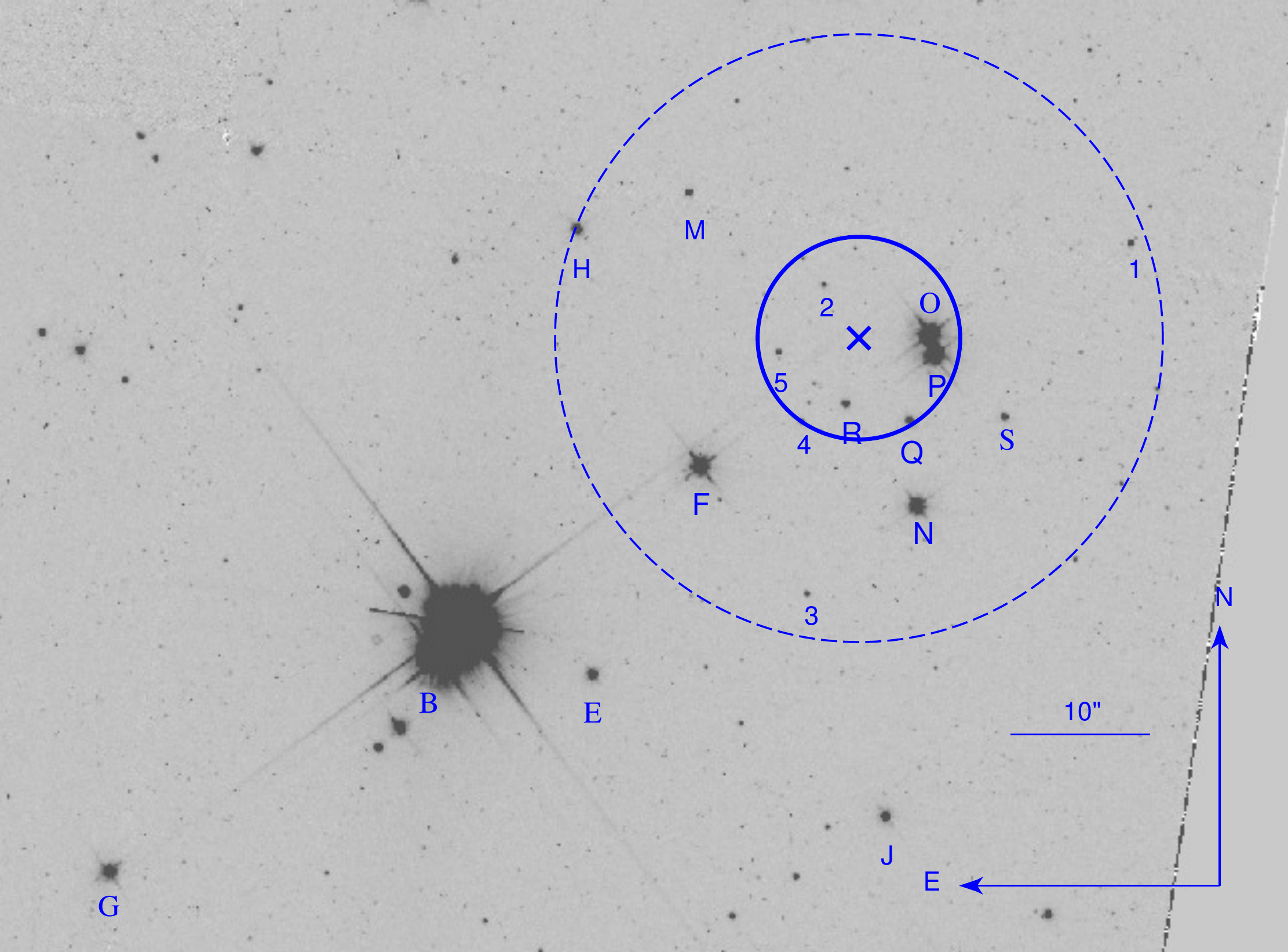}}	
         \caption{Final combined position for SN 1572.  Our combined position is marked with a blue X, the one-sigma error region is marked with the blue circle (7.3 arc-seconds in radius), while the three-sigma error region is marked by the thin blue dashed circle.  These are superposed on a 2810 second image through the F555W filter with the {\it Hubble Space Telescope} as taken on 1999-02-15.  This is one of the few {\it HST} images to show this region, with the edge of this image indicated by the nearly vertical line on the right edge of the image.  The labeling of stars with letters is from Bedin et al. (2014), and we have added numerical labels for other stars of interest.  About half of our final combined error region has not been examined for ex-companion candidates, and there is only this one epoch of minimal {\it HST} data.}
\end{figure}

\end{document}